\documentclass[conference]{IEEEtran}
\IEEEoverridecommandlockouts
\usepackage{cite}
\usepackage{amsmath,amssymb,amsfonts}
\usepackage{graphicx}
\usepackage{textcomp}

\usepackage{subfig}
\usepackage{algorithm, algpseudocode}
\usepackage{amsthm}
\newtheorem{thm}{Theorem}
\usepackage{multirow}

\renewcommand{\baselinestretch}{.9}
\begin{document}

\title{2D Proactive Uplink Resource Allocation Algorithm for Event Based MTC Applications}

\author{\IEEEauthorblockN{Thai~T.~Vu, Diep~N.~Nguyen, Eryk~Dutkiewicz}\IEEEauthorblockA{School of Computing and Communications, University of Technology,
		Sydney, NSW, Australia\\
		Email: TienThai.Vu@student.uts.edu.au, \{Diep.Nguyen, Eryk.Dutkiewicz\}@uts.edu.au}\IEEEauthorblockA{}}
	

\maketitle

\begin{abstract}
We propose a two dimension (2D) proactive uplink resource allocation (2D-PURA) algorithm that aims to reduce the delay/latency in event-based machine-type communications (MTC) applications. Specifically, when an event of interest occurs at a device, it tends to spread to the neighboring devices. Consequently, when a device has data to send to the base station (BS), its neighbors later are highly likely to transmit. Thus, we propose to cluster devices in the neighborhood around the event, also referred to as the disturbance region, into rings based on the distance from the original event. To reduce the uplink latency, we then proactively allocate resources for these rings. To evaluate the proposed algorithm, we analytically derive the mean uplink delay, the proportion of resource conservation due to successful allocations, and the proportion of uplink resource wastage due to unsuccessful allocations for 2D-PURA algorithm. Numerical results demonstrate that the proposed method can save over 16.5 and 27 percent of mean uplink delay, compared with the 1D algorithm and the standard method, respectively.
\end{abstract}

\begin{IEEEkeywords}
M2M, machine-type communications, proactive resource allocation, latency-sensitive services
\end{IEEEkeywords}

\section{Introduction}
Despite the original design of the cellular systems focusing on human-to-human (H2H) communications, the demands for machine-type communication (MTC), e.g., automation in industry 4.0, have created a new range of emerging services \cite{schulz2017latency}. In many MTC applications, such as monitoring systems, security and public safety services, a massive number of MTC devices are used to detect events in the environment, triggering uplink-centric transmissions of infrequent and low volume of delay-sensitive data\cite{shariatmadari2015machine}.

Various approaches at the MAC layer have been proposed to achieve the ultra-low delay in MTC. 
In \cite{wang2015optimal}, a scheme combining the access class barring (ACB) and timing advance information was proposed, thereby reducing half of random access slots necessary for MTC devices and improving the delay of both MTC and H2H networks. In \cite{wiriaatmadja2015hybrid}, a new hybrid protocol for random access and data transmission was developed to solve the random access channel congestion and excessive signaling overhead problems. Other methods are based on the traffic patterns of MTC and proactive resource allocations, e.g., \cite{wang2016internet,tadrous2015proactive, brown2015predictive}. Specifically, in \cite{brown2015predictive}, exploiting uplink traffic patterns between devices along a line, a proactive uplink resource allocation method, referred to as a 1D algorithm, is proposed for MTC. In this method, when a device sends a scheduling request (SR) for uplink resources, the BS can proactively grant uplink resources to a neighboring device along the line prior its next SR opportunity.
Here, an SR opportunity is a chance to send an SR at a certain time. 
Being allocated resource without sending an SR, the neighbor can reduce its uplink latency, compared with that of the LTE standard \cite{3GPP.ts36.321.v11.3.0}. However, the 1D algorithm is not suitable for practical scenarios, in which a large number of devices are deployed in a spatial region. In such scenarios, events or disturbances can occur at a position then spread to the surrounding area. 
Under the 1D algorithm, the BS considers a proactive resource allocation towards a device only when it receives an SR from another device.
Then, if the device does not satisfy conditions for a proactive resource allocation, it will follow the standard method by sending an SR. Therefore, at most 50 percent of
devices are proactively allocated resources while the rest, which follow the standard method, send SRs to the BS.


In this paper, 
we leverage the traffic correlations between devices in a spatial region to reduce their uplink latency. Specifically, devices in the disturbance area are clustered into rings \cite{lien2011toward} based on their distance to the event. Then, the BS targets those rings for proactive uplink resource allocations (i.e., prior their requests), thus reducing the uplink latency of all devices in the disturbance area.

We analyze the expected uplink delay and the proportions of resource conserved and wasted due to successful and unsuccessful predictions. The numerical results show that our method reduces more than 16.5 percent of the mean uplink delay, compared with the best case of the 1D algorithm. Major contributions of this paper are as follows:
\begin{itemize}
	\item We first propose a two dimension (2D) proactive uplink resource allocation (2D-PURA) algorithm, in which MTC devices in the disturbance region are spatially clustered into rings based on their distance to the original event. Then, these rings are proactively allocated resources for uplink transmissions. 
	\item We derive the uplink expected delay and the proportions of resource saving and wastage for 2D-PURA algorithm.
	\item Numerical results show that 2D-PURA outperforms the best case of the 1D algorithm \cite{brown2015predictive} and the standard method in LTE networks \cite{3GPP.ts36.321.v11.3.0} in terms of the mean uplink delay.
	\item We analyze the complexity and optimize the size of a ring of the 2D-PURA algorithm.
\end{itemize}

The rest of the paper is as follows. We describe the system model in Section \ref{sec:system-model}. Section \ref{sec:2D-Predictive} and \ref{sec:Mathematical-Analysis} present the proposed 2D-PURA algorithm and derive its performance characteristics. In Section \ref{sec:Simulation-and-Evaluation} we evaluate the performance of 2D-PURA algorithm and compare it with the 1D algorithm \cite{brown2015predictive} and the standard method \cite{3GPP.ts36.321.v11.3.0} previous ones. Conclusion is drawn in Section \ref{sec:Conclusion}.

\section{System Model}\label{sec:system-model}

Consider a spatial region in which MTC devices are randomly distributed according to a homogeneous Poisson Point Process (HPPP) \cite{tefek2017relaying} denoted by $\Phi$ with intensity $\lambda$. The BS or the IoT gateway that serves the region can proactively grant uplink resources to MTC devices instead of waiting for their requests for uplink transmissions. The location information of those devices is maintained at the BS. This assumption can be seen in many location-based wireless sensor networks (e.g., \cite{han2013localization} and therein references). Specifically, sensor devices can estimate their own location by an equipped GPS unit or using localization algorithms with the support of other GPS-enable devices such as smartphones. Then sensor nodes initially register their location information with the BS/gateway, e.g.,  \cite{zhu2015collaborative}, after being installed.

Exploiting the traffic correlations between devices to proactively allocate resources can be incorporated into any MTC, cellular, or IoT standard (e.g., NB-IoT). In this work, as a study case, we adopt the LTE standard. In LTE networks, a device can be in either RRC\_IDLE or RRC\_CONNECTED modes. In the former, the device can only receive broadcast information. Whilst in the latter mode, the device can send and receive data. The device must send an SR to request resources for uplink transmissions. 
For the ease of analysis,  we assume that all MTC devices have already been in the RRC\_CONNECTED mode \cite{3GPP.ts36.331.v14.1.0} of LTE networks with the Frequency Division Duplexing (FDD). 
After data enters the transmit buffer, devices must wait for their individual pre-assigned offset subframes (in LTE each subframe is 1 ms in duration) within an SR period $\sigma$ to send an SR to request an uplink grant. For simplicity, we assume throughout this paper that the pre-assigned offset follows a discrete uniform distribution on $\{1,2,..,\sigma\}$ denoted by $U(1,\sigma)$, and each device has a periodic SR opportunity every $\sigma$ subframes.

The expected standard uplink delay for an MTC device in LTE networks is
\begin{equation}
\label{eq_01}
E(D_{std})=(1+\sigma)/2+\beta+\delta
\end{equation}
where $(1+\sigma)/2$ subframes is the delay between the time instances the device has data in the transmit buffer and sends an SR. $\beta$ is the delay between the times the device sends an SR and receives an uplink grant from the BS (depending on the scheduling policy at the BS)\footnote{We assume the system is under a low load so that each uplink grant can occur at the earliest possibility or $\beta=1$.}. $\delta=4.5$ subframes include the average delay of 0.5 subframe that the data is available in the uplink buffer and the fixed delay of 4 subframes between the times the device receives an uplink grant and sends the data.

\begin{figure}[htbp]
\centerline{\includegraphics[scale=0.4]{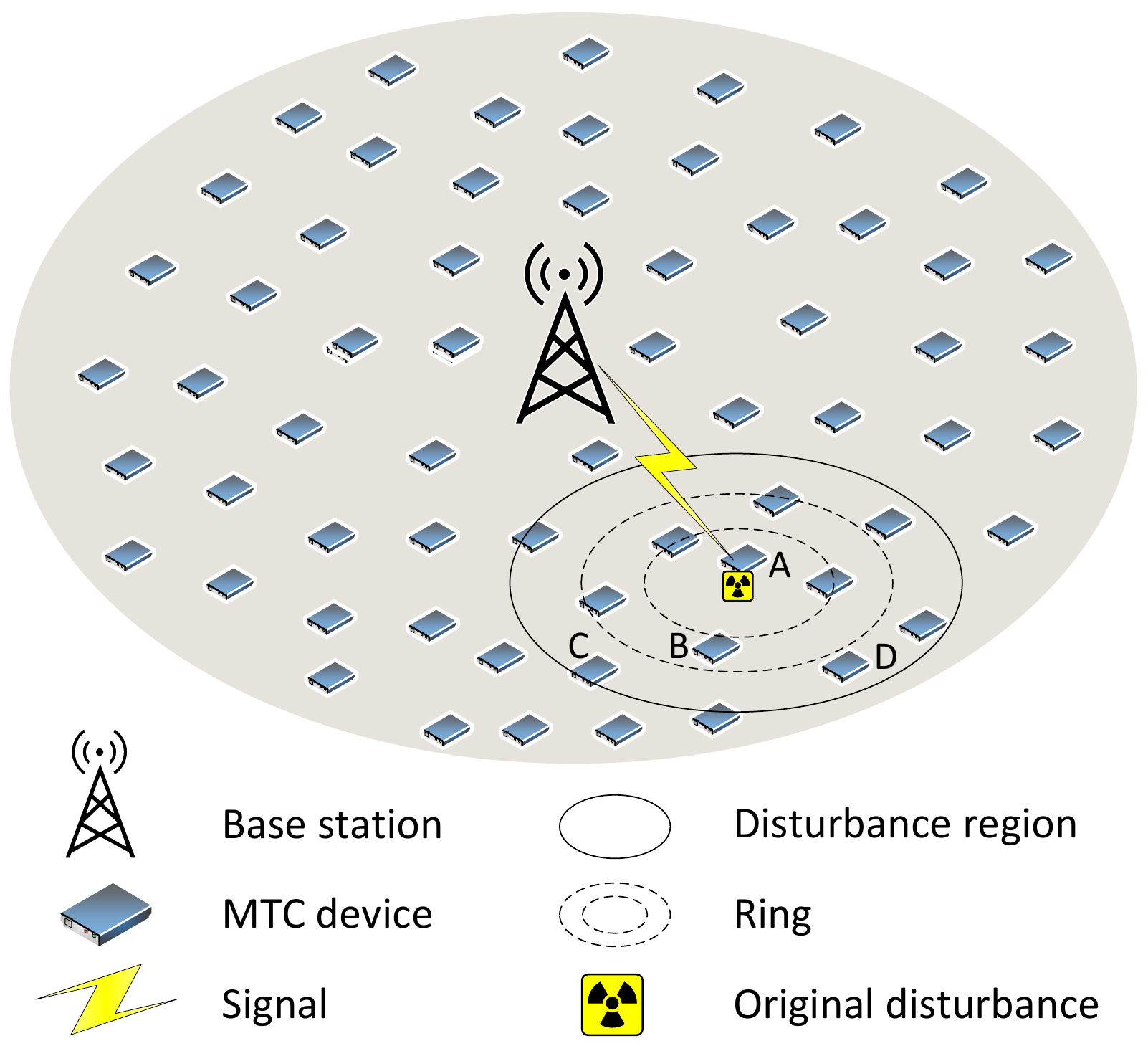}}
\caption{MTC communications under a cellular network model.}
\label{fig:Cell-Network-Model}
\end{figure}

This work aims to lower the mean uplink delay $E(D)$ by reducing the $(1+\sigma)/2$ component. 
To better demonstrate, in Figure \ref{fig:Cell-Network-Model}, when a disturbance initially occurs at device A, it later spreads and making a disturbance region. The BS clusters all devices in the disturbance region into rings based on their distance to device A, and then conducts proactive resource allocations for these rings in sequence from inside out. Specifically, when device A sends an SR to request uplink resource for pending data, later its neighbors such as device B, has a high probability of sending an SR. Hence, at the time receiving an SR from device A, the BS proactively allocates uplink resources for device B if
the time remaining to 
 its next SR opportunity is greater than a threshold $y$. Consequently, being granted uplink resources without sending an SR, those neighbor devices gain a lower uplink latency $E(D)$, compared with the reactive method.

We assume that the disturbance spreads with velocity $v$ to the surrounding area during a period $T$. Therefore, all devices in the disturbance region defined by a circle radius $ l=vT $ central device A can detect the same disturbance. In fact, each disturbance has different speed $v$ and a period $T$, and the BS does not know exactly this information. However, the BS can initially estimate the speed $ v $ and period $T$, based on the previous events (e.g. the average values of three latest disturbances), then, adjust these parameters for the current ring based on the results of the prediction for the previous ring.


\section{2D Proactive Uplink Resource Allocation Algorithm}\label{sec:2D-Predictive}
First, we present the proactive uplink resource allocation concept, then describe 2D-PURA algorithm for predicting devices on a spatial region.
\subsection{2D Proactive Uplink Resource Allocation}\label{sub:Predictive-Allocation-Concept}
Based on traffic patterns, the BS predicts the probability of a device having data in order to grant uplink resources in advance. Generally, when the BS receives an SR from device A, it can proactively allocate uplink resources for a neighbor device only when two following conditions are satisfied:
\begin{itemize}
	\item \textbf{Eligibility criterion} dictates that the neighbor device has not sent an SR recently, which is related to the same disturbance (i.e, the BS may configure the timer to determine the correlation between SRs).
	\item \textbf{Timing criterion} dictates that the time remaining to the next SR opportunity for the neighbor device is greater than threshold $y$ at the time the SR from device A is received.
\end{itemize}

Theoretically, all neighboring devices in the disturbance region can be applied the proactive resource allocation. However, with a massive number of MTC devices, it is impractical for the BS to calculate an individual threshold $y$ for each device, due to the resource constraint. 
On the other hand, if we use the same threshold $y$ for all devices in the disturbance region, the probability of successful predictions could drop dramatically because of the difference in their remaining time to the next SR opportunity.
\begin{figure}
	\centering
	\subfloat[][Clustering the disturbance\\ region into $n$ rings]{\includegraphics[width=4.4cm]{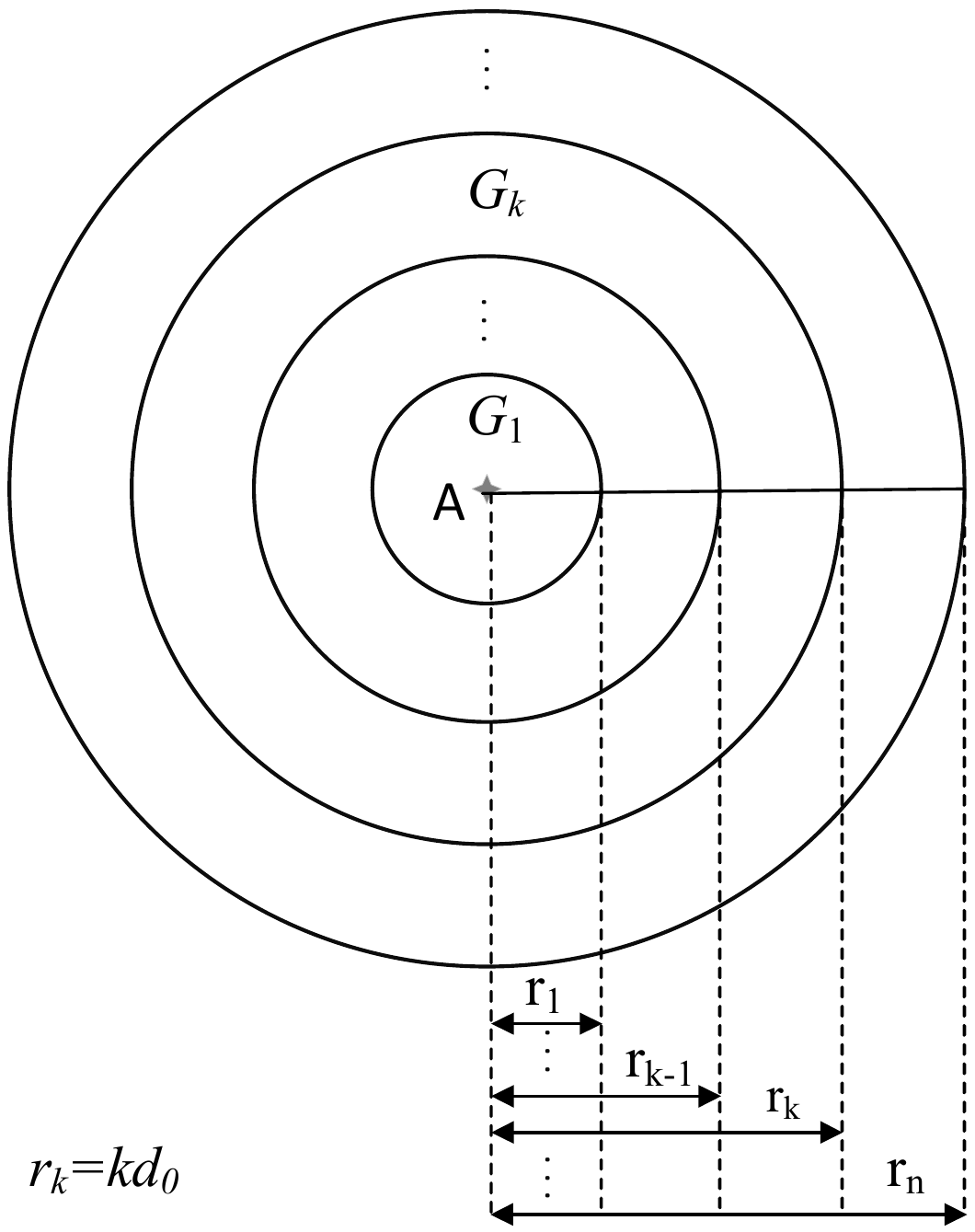}\label{sub.fig:Clustering-model}}
	\subfloat[][Predictive uplink resource\\ allocations for $k^{th}$ ring]{\includegraphics[width=4.4cm]{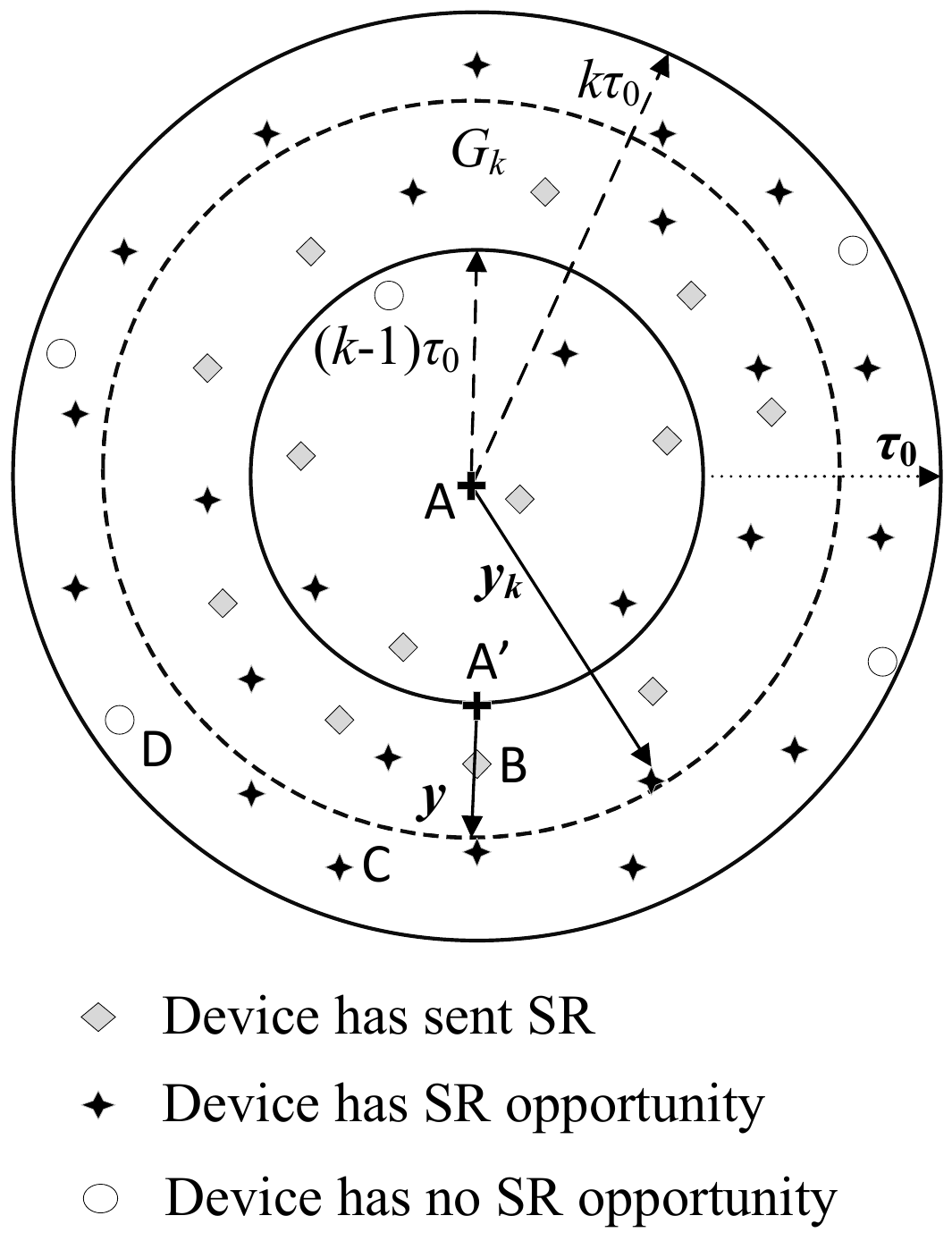}\label{sub.fig:Predictive-kth-group}}
	\caption{2D Predictive resource allocation model.}
	\label{fig:2D-Predictive-Model-2}
\end{figure}
To tackle it, in our 2D-PURA algorithm, all neighbor devices in the disturbance region are clustered into rings based upon their distance to event. The threshold $ y $, then, is used to proactively grant uplink resources one at a time to all devices in each ring, thereby reducing their average uplink delay.

We construct these rings with the same ring width $d_{0}$ as can be seen in Figure \ref{sub.fig:Clustering-model}. A set of circles center device A with radius $r_{k}=kd_{0}$ is used to cluster the neighbor devices into $n$ rings, where $n=\left\lceil l/d_{0}\right\rceil $, $l$ is the radius of the disturbance region. The $k^{th}$ ring is determined by the circle radius $r_{k-1}$ (the inner boundary) and
the circle radius $r_{k}$ (the outer boundary), and the necessary time for the disturbance to pass each ring is a constant and equals $\tau_{0}=d_{0}/v$, where $v$ is the speed of the disturbance.


The BS first exploits a standard uplink reactive resource allocation for device A since it is the first device informing about the pending data. Then, it proactively allocates uplink resources to all rings in sequence from inside out. To predict the $k^{th}$ ring based on the SR from device A, $y_{k}=(k-1)\tau_{0}+y$ is considered as a threshold. The component $(k-1)\tau_{0}$ is the necessary time for the disturbance to cross $(k-1)$ adjacent rings. In Figure \ref{sub.fig:Predictive-kth-group}, upon receiving an SR from device A, the BS starts the proactive uplink resource allocation for all devices in the $k^{th}$ ring. There are three possible cases at the time $y_{k}$ subframes after the SR from device A is received:

\begin{itemize}
	\item A device has already sent an SR (e.g., device B).
	\item A device has next SR opportunity (e.g., device C).
	\item A device has no SR opportunity (e.g., device D).
\end{itemize}

For device B, its next SR opportunity
occurs at less than
 $\ensuremath{y_{k}}$
subframes, then the BS does not proactively grant uplink resources
to device B, and a standard uplink resource allocation is applied.
For devices C and D, their remaining interval to the next SR opportunity
is greater than threshold $\ensuremath{y_{k}}$, then the BS proactively
allocates uplink resources to those neighbors prior to their next
SR opportunity. For device C, at the time of the proactive resource
allocation made, it has pending data, hence, the proactive resource
allocation is successful, thereby reducing its uplink latency. However,
when the proactive allocation occurs, device D has no pending data,
therefore, the proactive resource allocation is unsuccessful.

Here, we assume that the BS can set a timer so that it can detect the SR from device A is about the original disturbance instead of device B's. Therefore, device C and D will be proactively assigned uplink resources with respect to device A.

Of course,
our proposed method aims to reduce the delay of the whole network, thus
there is a risk of unsuccessful prediction in which the
BS allocates uplink resources to a device before it has data to send.
Threshold
$y$ is used to trade off uplink latency reduction
against resource wastage. We intensively analyze the resource wastage
both analytically and numerically in Sections \ref{sec:Mathematical-Analysis}
and \ref{sec:Simulation-and-Evaluation}.

\subsection{2D Predictive Uplink Resource Allocation Algorithm\label{sub:2D_PURA-Algorithm}}

	

\begin{algorithm}[tbh]
	\hspace{0.05cm} \textbf{Input}: Set of devices $S = \{D_1, D_2, \ldots, D_N\}$, device $A$
	
	\hspace{1.1cm} Spreading time of event $T$, speed of event $v$
	
	\hspace{1.1cm} Ring-crossing time of event $\tau_0$, threshold $y$
	
	\hspace{0.05cm} \textbf{Output}: Proactively allocating uplink resources to set $S$ 
	
	\begin{algorithmic}[1]
			\State $d_0 \gets v * \tau_0$
		\State $n \gets \lceil T*v/d_0 \rceil$
		\For{$k \gets 1, \ldots, n$}
		\State $G_k \gets \emptyset$
		\EndFor
		\ForAll {$D_i \in S$}	\Comment{Clustering $S$ into $n$ rings}
		\State $k \gets \lceil |D_i,A|/d_0 \rceil$
		\If {$k \leq n$}
		\State $G_k \gets G_k \cup \{D_i\}$
		\EndIf
		\EndFor
		\For{$k \gets 1, \ldots, n$}	\Comment{Proactively allocating uplink resources to $n$ rings}
		\State $y_k \gets (k-1)*\tau_0+y$
		\ForAll {$D_i \in G_k$}
		\If {(BS has not received SR from $D_i$) \textbf{and}
			
			\hspace{0.25cm} (Time to next SR opportunity for $D_i > y_k$)}
		\State BS prepares a proactive resource allocation for
		
		\hspace{0.7cm} $D_i$ to occur no  earlier than $y_k$
		\EndIf
		\EndFor	
		\EndFor
	\end{algorithmic}\caption{2D-PURA Algorithm($S$, $A$,  $T$, $v$, $\tau_{0}$, $y$)}
	\label{2D_PURA_code}
	
\end{algorithm}

The 2D-PURA algorithm's pseudocode 
is in Algorithm \ref{2D_PURA_code}. 
For device $D_{i}$ in the $k^{th}$ ring, we
assume that there is a virtual device $A'$, which is the intersection
of the inner boundary of the $k^{th}$ ring and the straight line
going through device A and $D_{i}$, for sending an SR about the disturbance
since it reaches the inner boundary as can be seen in Figure \ref{sub.fig:Predictive-kth-group}.
Therefore, device $D_{i}$ satisfies the \textit{timing} criterion (mentioned above)
if its next SR opportunity is $y$ subframes later than the time an
SR from the virtual device $A'$. Noticeably, the SR from the virtual
device $A'$ for the $k^{th}$ ring occurs $(k-1)\tau_{0}$ subframes after the SR from device A.
Therefore, in practice, the BS can proactively allocate uplink resources toward the $k^{th}$ ring at the time $y$ subframes after the SR from the virtual device $A'$ received. Devices also can send data to the BS while the algorithm is in process.

\subsection{Complexity Analysis}

When the BS receives an SR message and detects a new event, it runs 2D-PURA algorithm only one time in order to proactively allocate uplink resources to all MTC devices in the disturbance region. In 
Algorithm \ref{2D_PURA_code}, 2D-PURA has two stages, clustering and uplink resource allocations.
In the first stage, $N$ devices are clustered into $n$ rings with
the computational complexity is $O(N)$. The BS, then, proactively
allocates uplink resources to these $n$ rings. Every device in each ring is considered for a proactive uplink resource allocation. Thus, in the proactive allocation stage, the complexity is proportional to the number of devices and measured by $O(N)$. From both these stages, 2D-PURA has the complexity of $O(N)$.

\subsection{The Optimal Ring Width $d_{0}$}

The parameter $\tau_{0}=d_{0}/v$ characterizes the difference
in the next SR opportunities of devices in each ring. Since the
same threshold $y$ is used to predict all devices in each ring, the large value of $\tau_{0}$ leads to less accurate predictions, thereby increasing the expected uplink delay of all neighbor
devices in the disturbance region. Let $\tau_{max}$ be the maximum value of $\tau_{0}$ ($\tau_{max}\leq T$) satisfying the
expected uplink delay condition, so the maximum value of
$d_{0}$ is $\tau_{max}v$.

Let $N_{k}$ be the expected number of devices in the $k^{th}$ ring, then
$N_{k}=\lambda(\pi r_{k}^{2}-\pi r_{k-1}^{2})=\lambda\pi d_{0}^{2}(2k-1)$.
Therefore, the expected number of device in the largest ring (the $n^{th}$
ring) is $N_{n}=\lambda\pi d_{0}^{2}(2n-1)$. We have $n=\left\lceil \frac{l}{d_{0}}\right\rceil = \left\lceil \frac{vT}{v\tau_{0}}\right\rceil =\left\lceil \frac{T}{\tau_{0}}\right\rceil$. To simplify, we assume
that $\frac{T}{\tau_{0}}$ is an integer, so that $n=\frac{T}{\tau_{0}}$.
Consequently, we have $N_{n}=\lambda\pi v^{2}\tau_{0}(2T-\tau_{0})$.
Without loss of generality, assume that $2T\gg\tau_{0}$, thus $N_{n}\approx\lambda\pi v^{2}\tau_{0}2T$. Let $N_{max}$ is the maximum number of MTC devices that the BS can
grant uplink resources simultaneously in LTE-FDD networks. Ignoring
the probability of predictions targeting at the $n^{th}$ ring, we
have $N_{n}\leq N_{max}$, thereby $\tau_{0}\leq\frac{N_{max}}{\lambda\pi v^{2}2T}$.
The maximum possible value of $d_{0}$ satisfying the $N_{max}$ condition
is $\frac{N_{max}}{\lambda\pi v^{2}2T}v=\frac{N_{max}}{2\lambda\pi vT}$.

Choosing the width of a ring must be satisfied both the mean uplink
delay and $N_{max}$ conditions, therefore we have

\begin{equation}
d_{0}=min\left\{ \tau_{max}v,\frac{N_{max}}{2\lambda\pi vT}\right\} \label{eq:d_0_optimal-1}
\end{equation}

\section{Performance Analysis of 2D-PURA\label{sec:Mathematical-Analysis}}

In this section, we derive three performance metrics of 2D-PURA algorithm: the expected uplink delay $E(D)$,
the probability of saving SRs $P(SR)$ due to successful predictions,
and the probability of uplink resource wastage $P(W)$ due to unsuccessful
predictions. Let $P(S)$ and $P(U)$ be the probabilities of successful
and unsuccessful predictions. 

In 2D-PURA, if an unsuccessful proactive resource allocation happens to a device, it must send an SR to the BS to request uplink data resources. The probability of saving SRs over the disturbance region $\ensuremath{P(SR)}$ is given by $P\left(SR\right)=P(S)$. Similarly,
the probability of unsuccessful predictions expresses the uplink resource
wastage, thus, $P(W)=P(U)$.

For a device B in the disturbance region
radius $l=vT$, 
its location follows a uniform distribution in the disturbance
region with density $\rho=\frac{1}{\pi l^{2}}$ \cite{daley2002introduction}.

Consider the case where device B is at a specific position in the $k^{th}$ ring, in which $\tau$ is the necessary time for the disturbance
with speed $v$ to travel from a virtual device $A'$ on the inner
boundary to the device B. The BS targets device B for a proactive resource
allocation given that the BS has received an SR from the virtual device
$A'$. Let $\ensuremath{P(S_{B})}$ and $\ensuremath{P(U_{B})}$ respectively
be the probabilities of a successful/unsuccessful prediction targeting
at device B given that the BS has received an SR from the virtual
device $A'$. $E(D_{B})$ is the expected uplink delay of device B under this proactive resource allocation. Here, $\ensuremath{P(S_{B})}$,
$\ensuremath{P(U_{B})}$ and $E(D_{B})$ are functions of $\tau$.

\begin{thm}[]
	\label{thm:Density-theorem} The performance metrics, $E(D)$, $P(S)$
	and $P(U)$ are derived as
\begin{equation}
E(D)={\displaystyle \sum_{k=1}^{n}E^{(k)}}\left(D\right)
\end{equation}

\begin{equation}
P(S)={\displaystyle \sum_{k=1}^{n}P^{(k)}}\left(S\right)\label{eq:P_S_0-1}
\end{equation}

\begin{equation}
P(U)={\displaystyle \sum_{k=1}^{n}P^{(k)}}\left(U\right),
\end{equation}
where $\ensuremath{n=\left\lceil \frac{T}{\tau_{0}}\right\rceil }$,
and\\
\begin{equation}
E^{(k)}\left(D\right)=\frac{2}{T^{2}}\intop_{0}^{\tau_{0}}\left(\left(k-1\right)\tau_{0}+\tau\right)E(D_{B})d\tau,
\end{equation}
\begin{equation}
P^{(k)}\left(S\right)=\frac{2}{T^{2}}\intop_{0}^{\tau_{0}}\left(\left(k-1\right)\tau_{0}+\tau\right)P(S_{B})d\tau,
\end{equation}
\begin{equation}
P^{(k)}\left(U\right)=\frac{2}{T^{2}}\intop_{0}^{\tau_{0}}\left(\left(k-1\right)\tau_{0}+\tau\right)P(U_{B})d\tau.
\end{equation}
\end{thm}

\begin{IEEEproof}
	For the case of $E(D)$, let $r$ be the distance from device B to
	device A, then, the component of $E^{k}(D)$ of device B in the $k^{th}$
	ring with density $\rho$ is $E^{(k)}\left(D\right)=\intop_{r_{k-1}}^{r_{k}}\rho2\pi rE(D_{B})dr$.
	Applying $r=\left(\left(k-1\right)\tau_{0}+\tau\right)v$ and $\rho=\frac{1}{\pi v^{2}T^{2}}$
	into this expression, we have $E^{(k)}\left(D\right)=\frac{2}{T^{2}}\intop_{0}^{\tau_{0}}\left(\left(k-1\right)\tau_{0}+\tau\right)E(D_{B})d\tau$.
	For the special case of the $n^{th}$ ring, $E^{(n)}\left(D\right)=\frac{2}{T^{2}}\intop_{0}^{T-(n-1)\tau_{0}}\left(\left(k-1\right)\tau_{0}+\tau\right)E(D_{B})d\tau$.
	Since device B is uniformly distributed on the disturbance region,
	the measure $E(D)$ is the sum of components of the measure of device
	B in all rings, $E(D)={\displaystyle \sum_{k=1}^{n}E^{(k)}}\left(D\right)$.
	The cases of $P(S)$ and $P(U)$ are proved similarly.
\end{IEEEproof}

In \cite{brown2015predictive}, the authors restricted their predictive resource allocation problem to a sequence of devices along a line. Then, the idea of one-to-one proactive resource allocation was applied repeatedly for the sequence. In this idea, only one device is considered for a predictive allocation after the BS received an SR from another device. Consequently, at most 50 percent of devices is considered for a predictive resource allocation. In this work, we resolve a more general problem, in which devices are deployed on a two-dimensional plane. We then develop a one-to-many proactive resource allocation concept, when the BS receives an SR from device A, all neighboring devices will be considered for a proactive resource allocation. To improve the performance and effeciency, neighbour devices are clutered into rings based on the distance from device A, and the timing of proactive resource allocation targeting these rings will be deviated.

The one-to-one proactive resource allocation in \cite{brown2015predictive} is the basis of the one-to-many proactive resource allocation concept in our study. Therefore, we use the equivalent expressions of metrics $E(D_{B})$, $P(S_{B})$, and $P(U_{B})$ which were analyzed in \cite{brown2015predictive}:


\begin{equation}
E(D_{B})=\begin{cases}
\Gamma_{S1}(D_{B})+\Gamma_{U1}(D_{B})+\Gamma_{\overline{Pred}1}(D_{B}),  \quad y<\tau\\
\Gamma_{S2}(D_{B})+\Gamma_{\overline{Pred}2}(D_{B}), \quad \tau\leq y<\sigma
\end{cases}\label{eq:E_D_B_0}
\end{equation}

\begin{equation}
P(S_{B})=\begin{cases}
\frac{(\sigma-y-1)(\sigma+y)-(\tau-y-1)(\tau-y)}{2\sigma^{2}}, & y<\tau\\
\frac{(\sigma-y-1)(\sigma+2\tau-y)}{2\sigma^{2}}, & \tau\leq y<\sigma
\end{cases}\label{eq:P_S_B_0}
\end{equation}

\begin{equation}
P\left(U_{B}\right)=\begin{cases}
\frac{(\sigma-y-1)(\tau-y)}{\sigma^{2}}, & y<\tau\\
0, & \tau\leq y<\sigma
\end{cases}\label{eq:P_U_B_0}
\end{equation}
where $\tau$ is the necessary time for disturbance to move from the
virtual device $A'$ to device B, and

$\begin{aligned}\Gamma_{S1}(D_{B})= & \frac{y(\sigma-y-1)(y+2\delta+1)}{2\sigma^{2}}+\frac{\delta(\sigma-\tau)}{\sigma}\\
& +\frac{(\sigma-\delta)(\sigma-\tau)(\sigma+2y-\tau+1)}{2\sigma^{2}}\\
& -\frac{P_{\sigma+y-\tau}-P_{y}}{\sigma^{2}},
\end{aligned}
$

$\begin{aligned}\Gamma_{U1}(D_{B})= & \frac{(\tau-y)(\sigma-y-1)(\sigma-\tau+1)}{2\sigma^{2}}\\
& +\frac{(\tau-y)(\tau-y-1)\sigma}{2\sigma^{2}}\\
& +\frac{\left(\tau-y\right)\left(\sigma-y-1\right)\left(1+\delta\right)}{\sigma^{2}},
\end{aligned}
$

$\begin{aligned}\Gamma_{\overline{Pred}1}(D_{B})= & \frac{\left(y+1\right)\left(\tau-y\right)\left(2\sigma-\tau+3+2\delta\right)}{2\sigma^{2}}\\
& +\frac{y\left(y+1\right)\left(\sigma+3+2\delta\right)}{2\sigma^{2}}\\
& +\frac{\left(3+2\delta\right)\left(\sigma-\tau\right)\left(\sigma+2y-\tau+1\right)}{4\sigma^{2}}\\
& +\frac{P_{\sigma+y-\tau}-P_{y}}{2\sigma^{2}},
\end{aligned}
$

$\begin{aligned}\Gamma_{S2}(D_{B})= & \frac{\tau(\sigma-y-1)(2y-\tau+2\delta+1)}{2\sigma^{2}}+\frac{\delta(\sigma-y-1)}{\sigma}\\
& +\frac{(\sigma-\delta)(\sigma-y-1)(\sigma+y)}{2\sigma^{2}}-\frac{P_{\sigma-1}-P_{y}}{\sigma^{2}},
\end{aligned}
$

$\begin{aligned}\Gamma_{\overline{Pred}2}(D_{B})= & \frac{\tau\left(y+1\right)\left(y-\tau+3+2\delta\right)}{2\sigma^{2}}+\frac{\tau(\tau+1)}{2\sigma}\\
& +\frac{\left(3+2\delta\right)\left(\sigma-y-1\right)\left(\sigma+y\right)}{4\sigma^{2}}\\
& +\frac{P_{\sigma-1}-P_{y}}{2\sigma^{2}}+\frac{\left(y-\tau+1\right)\left(\sigma+3+2\delta\right)}{2\sigma}
\end{aligned}
.$

Here, $\ensuremath{P_{n}}$ is the $\ensuremath{n^{th}}$ square pyramidal
number given by $P_{n}={\displaystyle \sum_{k=1}^{n}k^{2}=\frac{n(n+1)(2n+1)}{6}}$.

\section{Numerical Results\label{sec:Simulation-and-Evaluation}}

In this section, we present numerical results to validate the performance
of 2D-PURA by comparing it with the standard method in LTE networks
\cite{3GPP.ts36.321.v11.3.0} and its equivalent 1D algorithm \cite{brown2015predictive}.

Since the final performance expressions are only related to time,
we describe the experiment scenarios by $\tau_{0}$ instead of $d_{0}$.
Details of parameters are given in the table below.

\begin{table}[htbp]
	\caption{Experimental parameters\label{tab:Experimental-parameters}}

	\centering{}%
	\begin{tabular}{|l|c|}
		\hline 
		Parameters & Value\tabularnewline
		\hline 
		Density $\lambda$ & 0.11\tabularnewline
		\hline 
		SR period $\sigma$ & 40ms\tabularnewline
		\hline 
		Threshold $y$  & 1-39ms\tabularnewline
		\hline 
		Disturbance time $T$ & 1000ms\tabularnewline
		\hline 
		Velocity of the disturbance & 300m/s\tabularnewline
		\hline 
		\multirow{1}{*}{Ring-crossing disturbance propagation time $\tau_{0}$} & 1-40ms\tabularnewline
		\hline 
	\end{tabular}
\end{table}

As mentioned in Section \ref{sec:system-model}, the standard uplink
latency in LTE networks is $E(D_{std})=(1+\sigma)/2+1+\delta$.
For the 1D algorithm on a spatial area, only when the BS receives an SR from a device, it considers a proactive resource allocation towards the nearest neighbor. Then, if this device does not satisfy the eligibility and timing criteria, it can not be targeted a proactive resource allocation and follows the standard method by sending an SR. Therefore, at most 50 percent of neighboring devices are proactively allocated resources while the rest send SRs to the BS following the standard method.
Noticeably, $\ensuremath{P(S_{B})}$, $\ensuremath{P(U_{B})}$
and $E(D_{B})$ are just performance metrics of a device since it is proactively granted uplink resources. Hence, in the best case, the mean uplink
delay is $E(D_{B}^{*})=\frac{(E(D_{std})+E(D_{B}))}{2}$, the probability
of saving SRs is $P(SR_{B}^{*})=\frac{P(S_{B})}{2}$, and the probability
of wasting uplink resources is $P(W_{B}^{*})=\frac{P(U_{B})}{2}$.

Let $E(X)$ be the average distance between two devices on a spatial area. When devices are deployed according to HPPP with density
$\ensuremath{\lambda}$, we have $E(X)=\frac{1}{2\sqrt{\lambda}}$ \cite{moltchanov2012distance}.
To numerically evaluate the 1D algorithm, we set $\tau=\tau_{avg}$ where $\tau_{avg}=\frac{E(X)}{v}$. 

With the speed of the disturbance $v$ and density $\lambda$ as
mentioned in Table~\ref{tab:Experimental-parameters}, the average
time required for the event to move between the two adjacent objects
is $\tau_{avg}=5$ms when $v=300$m/s. Here, the chosen speed
of the disturbance is approximately equal to the speed of sound
in air. While the parameter $\tau_{avg}$ for the 1D algorithm is a function
of $\lambda$ and $v$, the parameter $\tau_{0}$ for 2D-PURA can be chosen independently.
Therefore, 2D-PURA with different values of $\tau_{0}$ can be compared
with the 1D algorithm with $\tau=\tau_{avg}$ ($5$ms).

\begin{figure}[htbp]
	\centerline{\includegraphics[width=6.5cm]{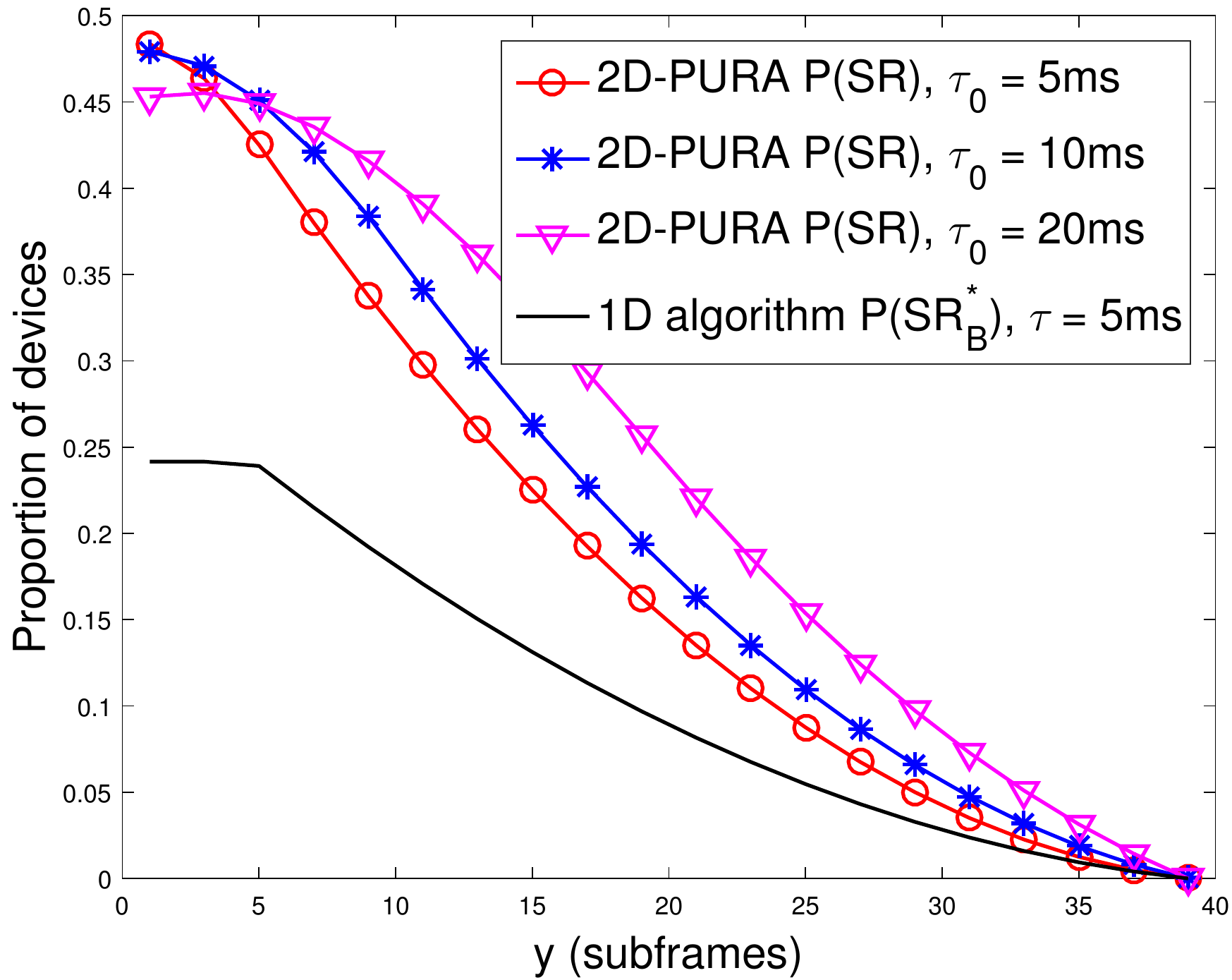}}
	\caption{SR saving due to successful predictions for 2D-PURA algorithm ($\sigma=40$
		subframes and various values of $\tau_{0}$).}
	\label{fig:SR-saving-y-vary}
\end{figure}

\begin{figure}[htbp]
	\centerline{\includegraphics[width=6.8cm]{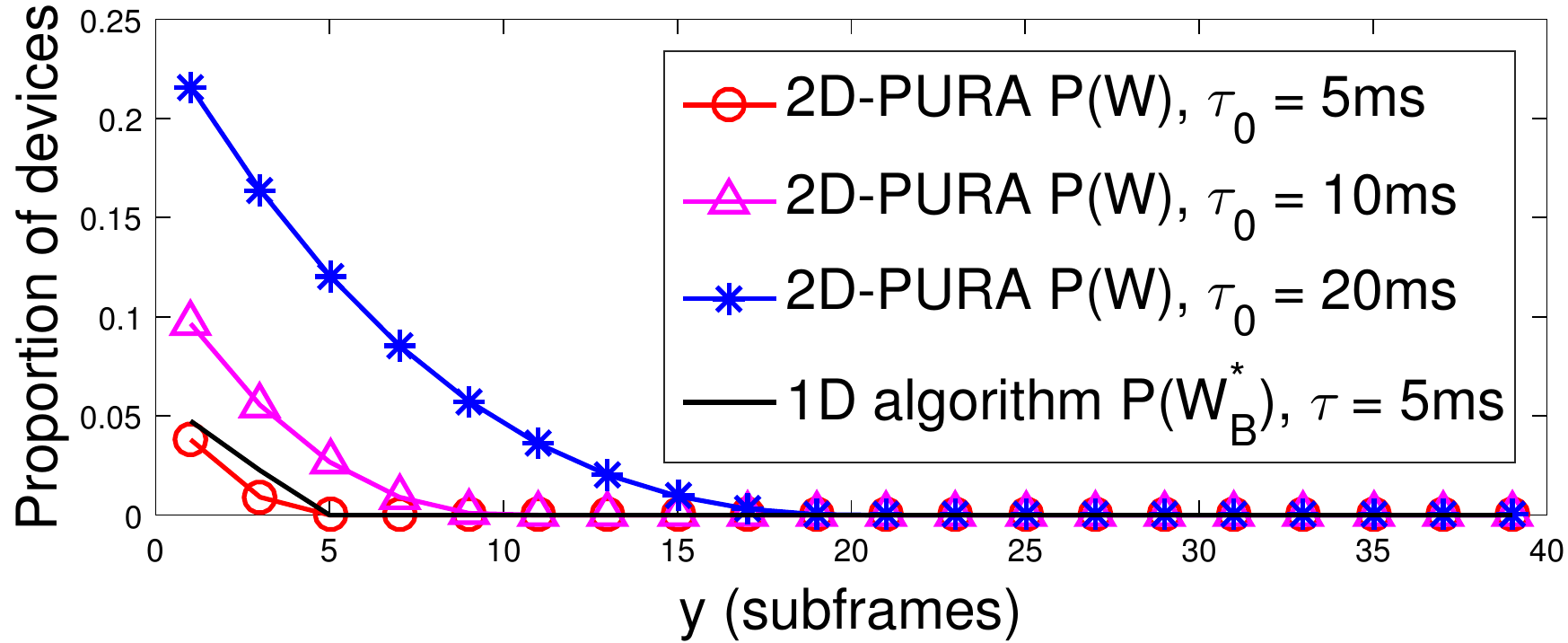}}
	\caption{Uplink resource wastage due to unsuccessful predictions for 2D-PURA
		algorithm ($\sigma=40$ subframes and various values of $\tau_{0}$).}
	\label{fig:Uplink-resource-wastage-y-vary}
\end{figure}

Figure \ref{fig:SR-saving-y-vary} depicts the proportion of devices
in the whole disturbance region transfering data without sending
SRs due to successful predictions. Generally, the downward trends
of SR saving are seen in both 2D-PURA and the 1D algorithms since
increasing the threshold $y$, and 2D-PURA with three different values
of parameter $\tau_{0}$ outperforms the 1D algorithm with $\tau=5$ms.
Especially, in the best case of both methods, 2D-PURA algorithm saves
$49$ percent of SR messages while 1D algorithm gains only 24 percent.
Figure \ref{fig:Uplink-resource-wastage-y-vary} shows the uplink
resource wastage in terms of the proportion of devices due to unsuccessful
predictions for both 2D-PURA and 1D algorithms. It is clear that the
uplink resource wastage decreases dramatically since increasing $y$,
and it reaches zero when $y\geq\tau_{0}$ for 2D-PURA and $y\geq\tau$
for the 1D algorithm. For 2D-PURA algorithm, the uplink resource wastage
is proportional to $\tau_{0}$ since $y<\tau_{0}$. For example, at
$y=1$ms, the uplink resource wastage of 2D-PURA algorithm for $\tau_{0}=20$ms
is twice that for $\tau_{0}=10$ms, and four times that for $\tau_{0}=5$ms
with the same value of $y\leq5$ms. Notably, due to the lower rate
being targeted proactive resource allocations, the uplink resource
wastage of 1D algorithm with $\tau=5$ms is much lower than that of
2D-PURA with $\tau_{0}=10$ms.

From Figures \ref{fig:SR-saving-y-vary} and \ref{fig:Uplink-resource-wastage-y-vary},
we can see that the more SRs we can save the more uplink resource
wastage we can suffer, and the threshold $y$ is used to trade off these metrics.

\begin{figure}[htbp]
	\centerline{\includegraphics[width=6.5cm]{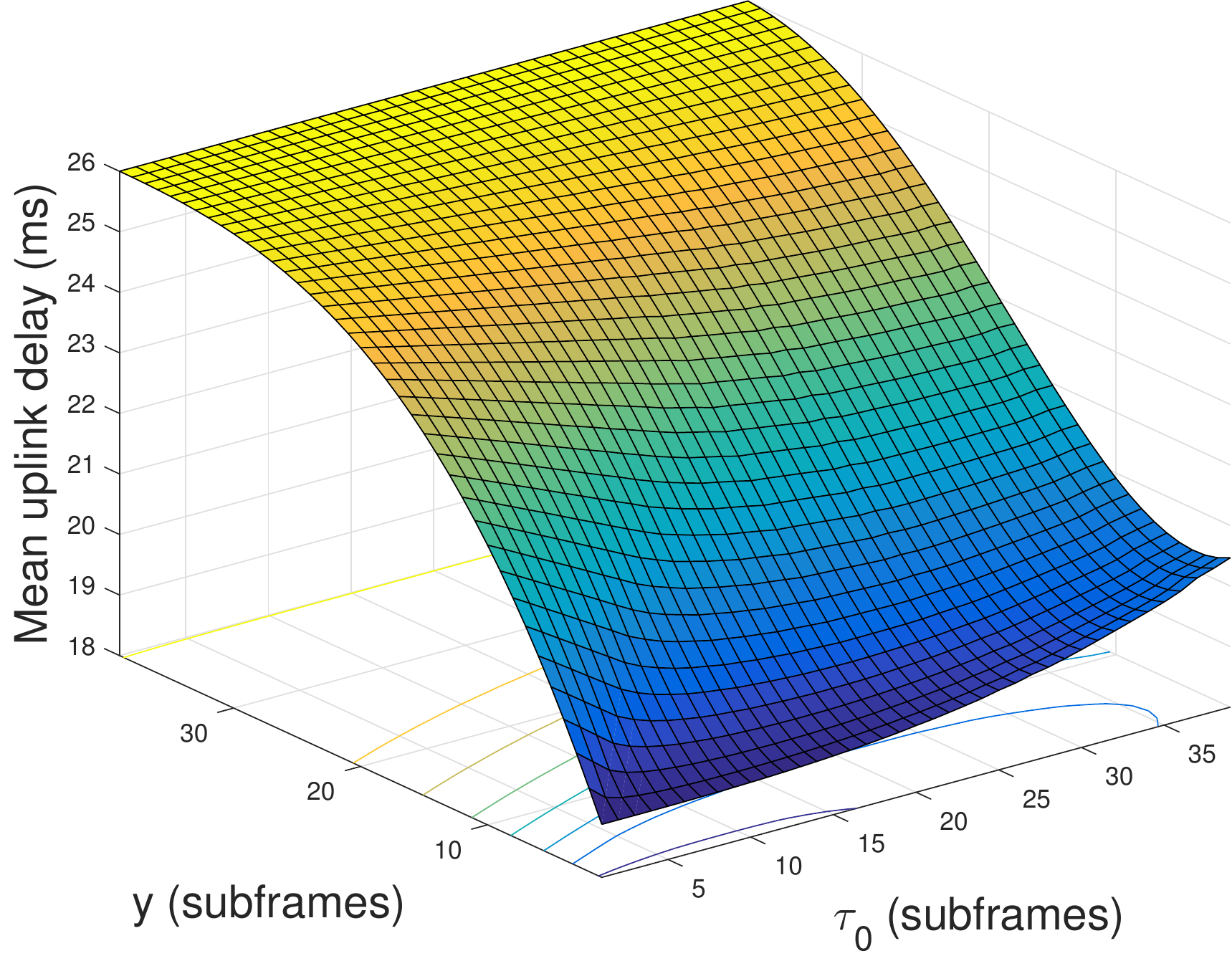}}
	\caption{Mean uplink delay for 2D-PURA algorithm ($\sigma=40$ subframes and
		various values of $\tau_{0}$ and $y$).}
	\label{fig:Mean-uplink-delay-y-tau-vary}
\end{figure}

To evaluate the effect of parameters $y$ and $\tau_{0}$ on the average
uplink latency of 2D-PURA algorithm, we conduct experiments with $y$
changing from $1$ms to $(\sigma-1)$ and $\tau_{0}$ changing from
$1$ms to $\sigma$ms. In Figure \ref{fig:Mean-uplink-delay-y-tau-vary},
for all values of $\tau_{0},$ the mean uplink delay generally increases
since $y$ rises from $1$ms to $(\sigma-1)$ms, and this metric gains
the minimum value, around $18.8$ms, when $y\leq2$ms and $\tau_{0}\leq20$ms.
This is because with the higher value of $y$, more devices, which
detect the disturbance early, are likely to send an SR prior the prediction
time, thereby increasing the mean uplink delay. Moreover, when increasing
$\tau_{0}$ from $1$ms to $\sigma$ms, the best values of the mean
uplink delay slightly raise from $18.8$ms to $20.4$ms. Thus, the
number of predictions can be reduced with a little cost of delay by
increasing the ring width of a ring in terms of time $\tau_{0}$.

\begin{figure}[htbp]
	\centerline{\includegraphics[width=6.5cm]{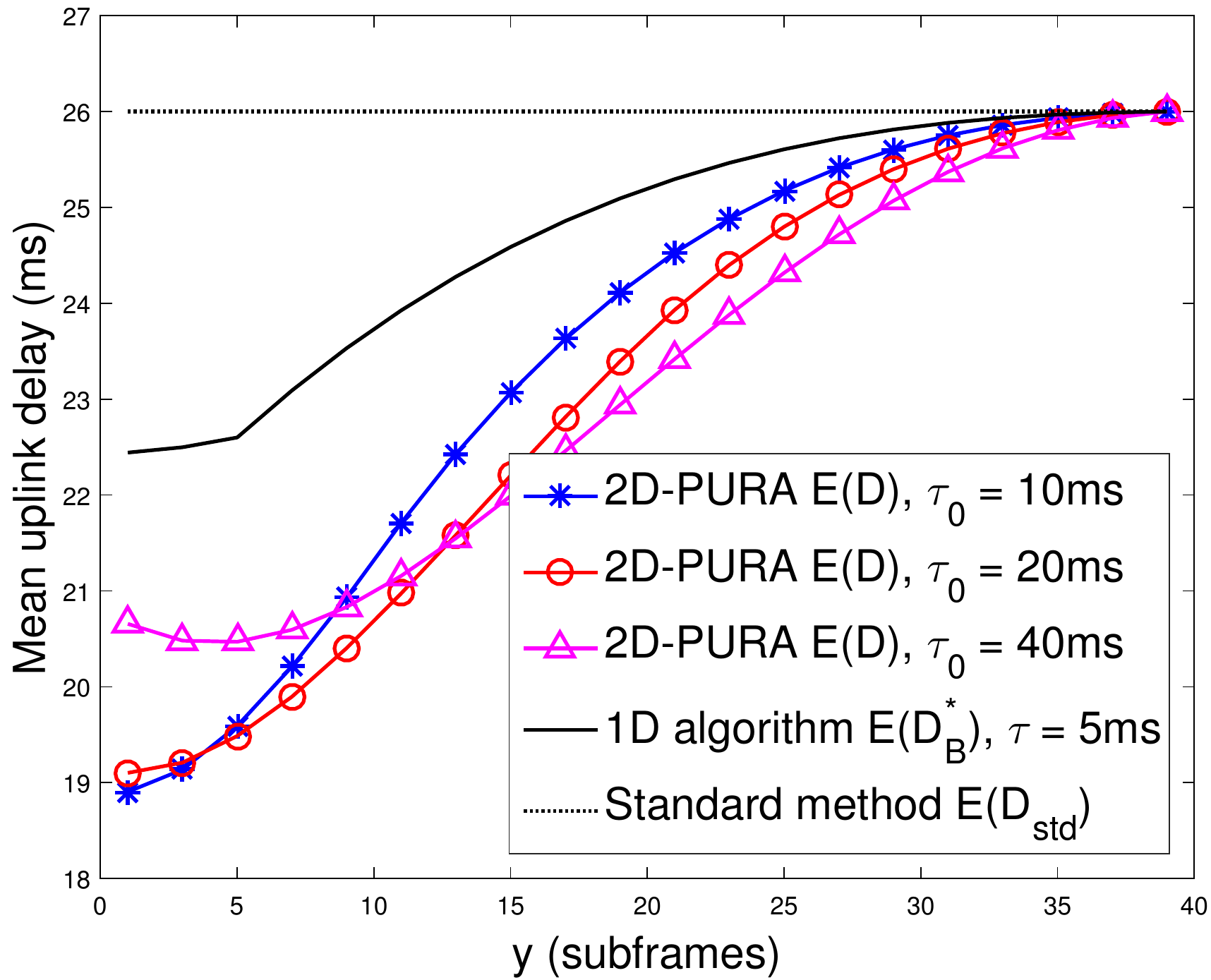}}
	\caption{A comparison of mean uplink delay for 2D-PURA, the 1D algorithm and
		the standard method ($\sigma=40$ subframes).}
	\label{fig:Mean-uplink-delay-y-vary}
\end{figure}

Figure \ref{fig:Mean-uplink-delay-y-vary} illustrates the mean uplink delay of 2D-PURA with different values of $\tau_{0}$, the 1D algorithm and the standard method. Generally, the expected delays of 2D-PURA and the 1D algorithm increase since the threshold $y$ goes up, and the figures reach that of the standard one. The best case of  the 1D algorithm $E(D_{B}^{*})$ with $\tau=5$ms on the whole disturbance region is much higher than that of 2D-PURA, $E(D)$, with all values of $\tau_{0}$. Specifically, when $y=1$ms 2D-PURA gains the mean uplink latency of $18.8$ms at $\tau_{0}=10$ms saving $27$ and $16.5$ percent consecutively in comparing with the standard method and the 1D algorithm. This is reasonable because 2D-PURA achieves a higher probability of proactive uplink resource allocations.
Thus, 2D-PURA algorithm not only resolves the proactive uplink
resource allocation on a spatial area but also outperforms the 1D
algorithm in terms of reduction of expected uplink delay.

In Figure \ref{fig:Mean-uplink-delay-y-vary}, the perfomance of 2D-PURA depends on both the values of $\tau_{0}$ and threshold $y$. In specific, the mean uplink delay of 2D-PURA with $\tau_{0}=10$ms is lower than that with $\tau_{0}=40$ms when the threshold $y < 10$ms and vice versa when $y \geq 10$ms. This is because with $10ms \leq y \leq 40$ms, the successful prediction rate of $\tau_{0}=10$ms (in this situation most of devices in the disturbance region send SR messages prior the time proactive allocations occur) is less than that of $\tau_{0}=40$ms. This completely matches with the dependance of the successful prediction rates on both $\tau_{0}$ and $y$ in Figure \ref{fig:SR-saving-y-vary}.

\section{Conclusion\label{sec:Conclusion}}

Exploiting the traffic patterns of MTC devices, we proposed a spatial or 2D proactive uplink resource allocation, called 2D-PURA. 
The algorithm significantly reduces the mean
uplink delay in comparison with the standard reactive method as well as the best case of the 1D algorithm (operating on a line of devices). We intensively investigated the proportions of resource conserved and wasted both analytically
and numerically. The complexity of 2D-PURA algorithm and the optimal ring width were also provided.

%

\end{document}